\documentstyle[aps,prd,preprint]{revtex}
\begin{document}
\draft

\title{Covariant Duality Symmetric Actions}
\author{A. Khoudeir\footnote[1]{adel@ciens.ula.ve} and N. R.
Pantoja\footnote[2]{pantoja@ciens.ula.ve}}
\address{\it Centro de Astrof\'{\i}sica Te\'orica,
Departamento de F\'{\i}sica, \\ Facultad de Ciencias,
Universidad de los Andes,\\ M\' erida 5101, Venezuela.\\}
\maketitle
\begin{abstract}
A manifestly Lorentz and diffeomorphism invariant form for
the abelian gauge field action with local duality symmetry
of Schwarz and Sen is given. Some of the underlying
symmetries of the covariant action are further considered.
The Noether conserved charge under continuous local duality
rotations is found. The covariant couplings with gravity
and the axidilaton field are discussed.
\end{abstract}
The equations of motion of the four dimensional low energy
effective field theory for the bosonic sector of the heterotic
string, which can be obtained from dimensional reduction of
$N=1$ supergravity theory coupled to gauge fields in ten
dimensions \cite{ms}, are invariant under the  $SL(2,R)$
non linear duality transformations of the massless fields
involved. This has been used to find new interesting
 black hole solutions carrying both electric and magnetic
charges in string theory \cite{shapere}. Actually,
$SL(2,Z)$ a subgroup of $SL(2,R)$, has been
conjectured to be an exact symmetry of the full string
 theory \cite{font}. This duality symmetry, called
S-duality, which also inverts the coupling constant,
 together with the "target space duality" or T-duality,
 have brought out new perspectives to the understanding
 of non perturbative features in string theory.

 Recently, Schwarz and Sen \cite{schwarz} have developed
 a method which permits to achieve $SL(2,R)$ duality
symmetry at the level of the action by introducing
extra auxiliary gauge fields. However, in their formulation,
 explicit Lorentz and general coordinate invariances are
missing. It is only after eliminating the auxiliary fields
through their equations of motion that the usual
transformation rules for the remaining fields in a specific
 gauge are recovered. In this paper we will consider the
manifestly Lorentz, gauge and diffeomorphism invariant
generalizations of the local duality symmetric actions
of Schwarz and Sen for abelian fields. Furthermore, some
of the underlying symmetries of the proposed actions are
discussed and the connection with previous works is
established.

 The simplest model of a duality symmetric action presented
 in Ref.\cite{schwarz}, deals with an appealing
generalization of the Maxwell electromagnetic theory in which
 two abelian gauge fields $A_{m}^{\alpha}$ ($\alpha = 1,2$)
are considered. Let us briefly discuss the main ideas. The
non-covariant action proposed by Schwarz and Sen is
\begin{equation}
I_{SS} = -\frac{1}{2} \int d^4x (B^{i\alpha}{\cal L}_
{\alpha \beta} E_{i}^{\beta} + B^{i\alpha}B^{\alpha}_i),
\label{actionss}
\end{equation}
where
\begin{equation}
E_{i}^{\alpha} = F_{oi}^{\alpha} = \partial_{o}A_{i}^{\alpha}
 - \partial_{i}A_{o}^{\alpha}, \quad
B^{i\alpha} = \frac {1}{2}\epsilon^{ijk}F_{jk}^{\alpha} =
\epsilon^{ijk}\partial_{j}A_{k}^{\alpha},
\end{equation}
and
\begin{equation}
{\cal L} = \left(
\begin{array}{cc}
 0 & 1\\
-1 & 0
\end{array}
\right) ,
\end{equation}
which has the following properties
\begin{equation}
det {\cal L} = 1, \quad
 {\cal L}^T = {\cal L}^{-1}, \quad
 {\cal S} {\cal L} {\cal S}^T = {\cal L},
\end{equation}
where ${\cal S}$ is a $SL(2,R)$ matrix. The action
is invariant under the following gauge
transformations
\begin{equation}
\delta A^{\alpha}_0 = \Psi^{\alpha}, \quad \delta A^{\alpha}_i=
\partial_i \Lambda^{\alpha}
\end{equation}
and the discrete duality transformations
\begin{equation}
A^{\alpha}_m \rightarrow {\cal L}_{\alpha\beta}A^{\beta}_m.
\label{duality}
\end{equation}
At the classical level, equivalence with the usual Maxwell
action in the temporal gauge $A^{\alpha}_0 = 0$ follows after
elimination of the fields $A^2_i$ using their equations of
motion. The duality transformation Eq. (\ref{duality}) reduces,
 on shell, to the well known transformation $\vec{E}
\rightarrow \vec{B}$ and $\vec{B} \rightarrow -\vec{E}$.
 It is worth recalling that the fields $A^{\alpha}_0$ do not
 play the usual role of Lagrange multipliers and that the
Gauss law constraint is consequence of $\partial_i
B^{\alpha i}= 0$. This equivalence holds also at the quantum
level as has been shown in Ref.\cite{martin}, where the
canonical quantization procedure reveals the presence of
second class constraints which may lead to serious problems
in supergravity models.

In the following we will consider the action
\begin{equation}
I= -\frac{1}{2}\int d^4x \ (u_n \ {\cal F}^{\alpha mn}
{\Phi}^{\alpha}_{mp}u^p + {\Lambda}^{\alpha mp}
{\Phi}^{\alpha}_{mp}),
\label{action}
\end{equation}
where
\begin{equation}
\Phi^{\alpha}_{mp} \equiv {\cal F}^{\alpha}_{mp} +
{\cal L}_{\alpha\beta}F^{\beta}_{pm},
\end{equation}
$u$ is an otherwise arbitrary vector field that satisfies
\begin{equation}
u_{m} u^m=-1,
\label{norm}
\end{equation}
$F$ and ${\cal F}$ are the gauge invariant field strengths
and their duals
\begin{equation}
F^{\alpha}_{mn}= \partial_mA^{\alpha}_n-\partial_nA^{\alpha}_m,
\label{F} \quad
{\cal F}^{\alpha mn}= \frac{1}{2} \varepsilon^{mnpq}
F^{\alpha}_{pq},
\label{dual}
\end{equation}
and ${\Lambda}$ is an auxiliary antisymmetric field. Note that
${\Phi}$ is a self-dual tensor, i.e.,
\begin{equation}
\Phi^{\alpha}_{mn} \equiv \frac{1}{2} \varepsilon^{mnpq}
{\cal L}_{\alpha
\beta}{\Phi}^{\beta pq}
\end{equation}
and hence ${\Lambda}$ is an anti self-dual tensor
\begin{equation}
{\Lambda}^{\alpha}_{mn} \equiv -\frac{1}{2} \varepsilon^{mnpq}
{\cal L}_{\alpha \beta}{\Lambda}^{\beta pq}.
\end{equation}
 Our conventions are $\eta_{mn}= diag(-1,1,1,1)$,
$\varepsilon^{ijk} = \varepsilon^{0ijk}$ and
$\varepsilon^{0123}= 1$. Clearly, the action Eq.(\ref{action})
is Lorentz and gauge invariant and manifestly invariant under
the duality transformations provided ${\Lambda}$ transforms as
follows:
\begin{equation}
{\Lambda}^{\alpha}_{mn} \rightarrow {\cal L}_{\alpha\beta}
{\Lambda}^{\beta}_{mn}.
\end{equation}

The equations of motion obtained from the action Eq.(\ref{action})
are:
\begin{equation}
\frac{\delta}{\delta A^{\alpha}_m} I = 0 \quad \Rightarrow \quad
G^{\alpha m} \equiv \varepsilon^{mnpq}\partial_p
[ u_n u^r {\Phi}^{\alpha}_{qr} - {\Lambda}^{\alpha}_{nq}] = 0,
\label{E-L}
\end{equation}
\begin{equation}
\frac{\delta}{\delta u_n} I = 0 \quad \Rightarrow \quad
H_{n} \equiv  \ {\cal F}^{\alpha}_{mn}{\Phi}^{\alpha mp}u_p +
{\cal F}^{\alpha}_{mp}{\Phi}^{\alpha mn}u^p= 0,
\end{equation}
and
\begin{equation}
\frac{\delta}{\delta \Lambda^{\alpha}_{mn}} I = 0
\quad \Rightarrow \quad {\Phi}^{\alpha}_{mn} = 0.
\label{link}
\end{equation}
Note that ${\partial}_m {\Phi}^{\alpha mn} = 0$ implies
${\partial}_m F^{\alpha mn} = 0$ which are just the Maxwell
equations for the $A^{\alpha}$ fields. One can show that
${\Lambda} = 0$ on-shell, i.e. is a non-dynamical variable.
After eliminating one of the two abelian gauge fields
(for example $A^2$) with the use of Eq. (\ref{link}) and
taking into account that $u_{n}u^{p}{\cal F}^{1mn} F^{2}_{pm}
 = -u_{n}u^{p}{\cal F}^{2mn} F^{1}_{pm}$ up to a total
divergence we obtain the gauge invariant Maxwell covariant
action $I= -\frac{1}{4} \int d^4x F^1_{mn}F^{1mn}$ for
the other gauge field.

Contact with the non-covariant formulation of Schwarz and
Sen can be made as follows. Our action has an additional
gauge symmetry:
\begin{eqnarray}
{\delta}u_m = {\epsilon}_m(x),
\quad {\delta}{\Lambda}^{\alpha}_{mn} = -\frac{1}{2}
[ {\epsilon}^r(x) {\cal F}^{\alpha}_{mr}u_n -
u^r{\cal F}^{\alpha}_{mr}{\epsilon}_n(x) -
(m \leftrightarrow n) ],
\end{eqnarray}
while the vector fields $A^{\alpha}_{m}$ are inert under
these transformations. This symmetry
allows one to fix the vector field $u$ to a constant vector,
e.g. $u_m = -\delta^0_m$. With this choice
 the action of Schwarz and Sen is obtained.

The action Eq. (\ref{action}) is not only invariant under
the discrete duality transformations; actually, it is
invariant under the continuous duality rotations
\begin{equation}
A^{\alpha m} \rightarrow {\cal D}_{\alpha\beta}A^{\beta m},
\label{dualrot}
\end{equation}
where ${\cal D}$ is a $SO(2,R)$ matrix
\begin{equation}
 {\cal D} = \left(
\begin{array}{cc}
 cos \omega & sin \omega\\
-sin \omega & cos \omega
\end{array}
\right),
\end{equation}
which implies the existence of conserved currents. If we
consider the infinitesimal duality rotations
\begin{equation}
\delta A^{\alpha}_m = \delta \omega {\cal L}_{\alpha \beta}
A^{\beta}_m,
\end{equation}
the Noether conserved current associated to this invariance is
\begin{equation}
j^m= -\frac{1}{2} \ {\cal F}^{\alpha mn}A^{\alpha}_ n,
\label{current}
\end{equation}
which satisfies $\partial_m j^m=0$ on shell.This current is
not gauge invariant, but it changes under gauge transformations
 by the divergence of an antisymmetric tensor. The corresponding
 conserved generator is
\begin{equation}
G= - \frac{1}{4} \int d^3x \varepsilon^{ijk} F^{\alpha}_{jk}
A^{\alpha}_i,
 \label{generator}
\end{equation}
which can be rewritten as
\begin{equation}
G= \int d^3x A^{\alpha}_i {\cal L}_{\alpha \beta} \Pi^{\beta i},
\end{equation}
where $ \Pi^{\beta i}= -\frac{1}{2} {\cal L}_{\alpha \beta}
B^{i \alpha} $ are the canonical conjugate momenta.
The integrand in Eq. (\ref{generator}) is just the sum of two
abelian Chern-Simons 3-form and $G$ as given by this equation
 is gauge invariant. Note that, in spite of the topological
 nature of the Chern-Simons forms, the generator
Eq. (\ref{generator}) is a genuine Noether charge. Hence the
 Chern-Simons 3-forms constructed out with the fiels
$A^{\alpha}_i$ are the generators of the continuous duality
 rotations, i.e., $\delta A^{\alpha}_i= \delta \omega
\{G,A^{\alpha}_i\}$. At this stage, Eq. (\ref{link}) can be
used to eliminate $ A^2_m $ from Eq. (\ref{generator}) to obtain
\begin{equation}
G= \frac{1}{2} \int d^3x (-\vec{A} \cdot \nabla \times \vec{A} +
\vec{E} \cdot \nabla^{-2} \nabla \times \vec{E}),
\end{equation}
which turns out to be the generator of the non-local duality
transformations in the usual canonical Maxwell theory
\cite{deser}.

Starting out with the covariant action the
 coupling with gravity keeping the duality invariance is now
straighforward. Following the minimal coupling prescription we
have
\begin{equation}
I_c= -\frac{1}{2}\int d^4x \sqrt{-g} \ (u_n
{\cal F}^{\alpha mn}{\Phi}^{\beta}_{pm}g^{pq}u_q +
g^{mn}g^{pq}{\Lambda}^{\alpha}_{ mp}{\Phi}^{\alpha}_{nq}),
\label{curved-action}
\end{equation}
where now
\begin{equation}
{\cal F}^{\alpha mn}= \frac{1}{2\sqrt{-g}} \varepsilon^{mnpq}
F^{\alpha}_{pq}.
\end{equation}
This action is manifestly invariant under general coordinate
 transformations and manifestly gauge and duality rotations
invariant. It is also invariant under conformal transformations
\begin{equation}
g_{mn} \rightarrow \Omega^2 g_{mn}, \quad F_{mn}^{\alpha},
\rightarrow F_{mn}^{\alpha}, \quad {\Lambda}^{\alpha}_{mn}
\rightarrow {\Lambda}_{mn}^{\alpha}
\end{equation}
provided
\begin{equation}
u_m \rightarrow \Omega u_m.
\end{equation}
The constraint imposed by ${\Lambda}$ turns out to be the same
 as in Eq. (\ref{link}). This can be used to eliminate the
field $A^2_m$ and the covariant Maxwell action in curved
space-time is obtained.
 Furthermore, a traceless on shell energy-momentum tensor
can be derived through the usual definition $T^{mn} \equiv
\frac{2}{\sqrt{-g}} \frac{\delta I_c}{\delta g_{mn}}$,
in agreement with the conformal invariance of the action.
>From this and after eliminating the field $A^2_m$ using
Eq. (\ref{link}), the Maxwell energy-momentum tensor is
obtained. For $u_m = -n \delta^0_m$ it
follows from Eq. (\ref{norm}) that $n^2= (-g^{00})^{-1}$
and contact with the usual ADM slicing of the manifold
\cite{arnowitt} is made. In this way, the local duality
invariant action of Schwarz and Sen in curved-space
\cite{schwarz}is recovered.

Finally, let us introduce the complex scalar axidilaton
field $\lambda \equiv\lambda_1 + i \lambda_2$, which
combines two entities of stringy nature: the dilaton
$\lambda_2 = exp \{-2\phi\}$ and the axion $\lambda_1
= \psi$ defined through the equations of motion of the
antisymmetric two-form $B_{mn}$ in four dimensions.
Introducing the symmetric $SL(2,R)$ matrix \cite{schwarz}
\begin{equation}
 {\cal M} = \frac{1}{\lambda_2} \left(
\begin{array}{cc}
 1 & {\lambda}_1\\
{\lambda}_1 & |{\lambda}|^2
\end{array}
\right) ,
\end{equation}
the covariant action for the coupling between the axidilaton
 and the gauge fields $A_m^{\alpha}$ in a curved space is
\begin{equation}
\begin{array}{c}
I_{ca}= -\frac{1}{2}\int d^4x \sqrt{-g} \ [u_n
{\cal F}^{\alpha mn}
( {\cal L}_{\alpha\beta} F^{\beta}_{pm} + \ ({\cal L}^T {\cal M}
{\cal L})_{\alpha \beta}{\cal F}^{\beta}_{mp})g^{pq}u_q \\+
g^{mn}g^{pq}{\Lambda}^{\alpha}_{ mp}( {\cal L}_{\alpha\beta}
F^{\beta}_{nq} + \ ({\cal L}^T {\cal M}
{\cal L})_{\alpha \beta}{\cal F}^{\beta}_{qn})],
\label{curved-axidilaton}
\end{array}
\end{equation}
which is invariant under the $SL(2,R)$ transformations
\begin{equation}
{\cal M} \rightarrow {w} {\cal M} {w}^T, \quad
A^{\alpha}_m \rightarrow (w)_{\alpha \beta} A^{\beta}_m.
\end{equation}
In fact, after eliminating the field $A^2_m$ through the
corresponding duality condition which arise from the constraint
 imposed by ${\Lambda}$
\begin{equation}
{\cal F}^{2mn}= \lambda_2 F^{1nm} + \lambda_1 {\cal F}^{1mn},
\end{equation}
the result is
\begin{equation}
I_{ca}= -\frac{1}{4}\int d^4x \sqrt{-g} \ F^1_{mn}
( \lambda_2 F^{1mn} + \ \lambda_1 {\cal F}^{1mn}),
\end{equation}
which exhibits the standard coupling of the axidilaton field
with an abelian gauge field in a curved space.

Summarizing, introducing auxiliary fields, we have seen
 that local duality transformations in the sense of Schwarz
and Sen can be implemented in a manifestly Lorentz invariant
way as a symmetry of the action, the field ${\Lambda}$ playing
the role of a multiplier whose associated constraint turns
out to be the covariant duality condition. The equivalence
with the covariant Maxwell theory follows after solving this
constraint in order to eliminate one of the gauge fields
from the original action. The connection with the non-covariant
approach is also established as a result of an additional
symmetry which permits to fix $u$ appropriately.
The generator for the continuous duality
transformations was found to be given in terms of
Chern-Simons 3-forms. From the proposed covariant action,
the coupling with gravity was obtained in a straighforward
way. In addition, the presence of the axidilaton field
coupled with the abelian gauge fields in a curved space
was considered.

We kindly thank Hector Rago, Umberto Percoco and Alvaro
Restuccia for fruitful discussions and constant interest
 in this work. This paper is dedicated to the memory of
Professor Carlos Aragone.

\end{document}